\documentclass[twocolumn,showpacs,preprintnumbers,amsmath,amssymb]{revtex4}

\usepackage{graphicx}
\usepackage{dcolumn}
\usepackage{bm}
\usepackage{xspace} 

\newcommand{\ket}[1]{|{#1}\rangle}
 
 \newcommand{\kbold}{\mathbf{k}}
 \def\mum{$\mu$m}

\def\kbold{\mathbf{k}}
\def\rjbold{\mathbf{r}_j}
\def\rjpbold{\mathbf{r}_{j'}}

\def\Rb87{$^{87}\text{Rb}$}
\def\0{\ket{0}}
\def\1{\ket{1}}

\begin{document}

\preprint{APS/123-QED}

\title{Few qubit atom-light interfaces with collective encoding}

\author{Line Hjortsh\o j Pedersen and Klaus M\o lmer}
\affiliation{
Lundbeck Foundation Theoretical Center for Quantum System Research, Department of
Physics and Astronomy, University of Aarhus, DK-8000 \AA rhus C, Denmark.}

\date{\today}

\begin{abstract}

Samples with few hundred atoms within a few $\mu$m sized region of space are large enough to provide efficient cooperative absorption and emission of light, and small enough to ensure strong dipole-dipole interactions when atoms are excited into high-lying Rydberg states. Based on a recently proposed collective encoding scheme, we propose to build few-qubit quantum registers in such samples. The registers can store and transmit quantum information in the form of single photons, and they can employ entanglement pumping protocols to perform ideally in networks for scalable quantum computing and long distance quantum communication.


\end{abstract}

\maketitle

Physicists are currently exploring the potential of quantum
information processing with particular efforts devoted to the
construction of practical devices for quantum communication and
quantum computing. The most challenging task for quantum
communication is to reach long distances, and for quantum computing
it is to achieve scalability towards a significant number of
controllable quantum bits. A combination of stationary qubits in
atoms or other material encoding media and flying qubits in photons,
transmitted along wave guides, through optical fibres or free space,
is expected to be an important ingredient in the achievement of the
ambitious long term goals for quantum information, and interfacing
of stationary and flying qubits constitutes a very active field of
research. Following the proposal for quantum repeaters based on
optically dense media \cite{dlcz}, significant progress has been
obtained \cite{kimble,kuzmich} on the entanglement of collective
qubit degrees of freedom in physically separated atomic samples.
These experiments involve large atomic gases, and while errors and
imperfections may be detected and "repeat-until-successful"
strategies may be employed to secure the formation of entangled
states \cite{dlcz,schmied}, effective restoration of quantum
information by error correction procedures is difficult to achieve
in these systems. The latter problem would be solved by interfacing
light with a small quantum processor, allowing one- and two-bit
gates among its qubits. Interfacing light with a single atom or ion
can be done with probabilistic protocols
\cite{duankimble03,childress35,monroe,rempe}, and hence one can
entangle remote quantum processors in a conditional manner
\cite{monroe2}. Although progress has been made recently on the free
space coupling of a single atom to a focussed single photon field
\cite{leuchs,kurtsiefer}, light couples more strongly and with a
larger degree of directional selectivity to an atomic ensemble, and
we shall show in the following that high probability heralded
schemes and deterministic schemes are within reach for the coupling
of photonic qubits to a small atomic ensemble of few hundred atoms.
In contrast to the macroscopic samples \cite{kimble,kuzmich}, we
suggest to confine the atoms within a 10 $\mu$m wide volume so that
the Rydberg blockade mechanism \cite{jaksch} can be used for quantum
gate operations on collectively encoded qubits \cite{coll1,coll2} in
different internal states in the atomic ensembles. It was recently proposed
\cite{petr} to extend the Rydberg blockade to macroscopic distances by use of a
stripline cavity field and thus obtain an efficient light-atom interface.
Our multi-qubit collective encoding scheme may also apply in that system, but we shall here focus
on the smaller samples.

A collection of atoms exhibits cooperative spontaneous emission
\cite{D54} and following \cite{lukinens},  it was
shown in \cite{SW02} that even a fairly small cloud of Rydberg blocked atoms
constitutes a directional source of single photons. Our analysis is
based on a solution of the full time-dependent problem of light
emission, as we aim to extract precise information about the
spatio-temporal field mode coupled to our system. We consider first
the case of emission, which may be used as a heralding event in
repeater and small register networks, and we then use our results to
assess how large a fidelity may be obtained in a direct writing of
photonic information on our stationary qubit in a deterministic
protocol.

We assume the initial collective atomic state
\begin{equation}\label{eq:psi0}
\ket{\Psi_0} = \frac{1}{\sqrt{N}}\sum_{j=1}^N e^{i\mathbf{k}_0 \cdot
\mathbf{r}_j} \ket{e_j}\otimes \ket{0},
\end{equation}
where $\ket{e_j}$ is shorthand for the state with atom $j$ excited and the
other atoms in their ground state $|g\rangle$. This state may be prepared using laser fields with wave vectors  $\mathbf{k}_{1,2}$ driving a resonant two-photon excitation into a Rydberg state, so that the blockade interaction prevents transfer of more than a single atom to the Rydberg excited state. A resonant $\pi$-pulse with wave vector $\mathbf{k}_3$ hereafter drives the atomic excitation into the excited state $|e\rangle$, producing the state (\ref{eq:psi0}) with  $\mathbf{k}_0=\mathbf{k}_1+\mathbf{k}_2-\mathbf{k}_3$.

To study the time dependence of the emission of light from the
excited ensemble, we shall use the approach in \cite{MK07}. For
simplicity, effects of photon polarization are neglected, but may
readily be incorporated in a more detailed analysis. The atoms and
the quantized field modes are governed by the Hamiltonian
$H_0  = \sum_{j=1}^N \hbar\omega_0 |e_j\rangle\langle e_j| +
\sum_{\mathbf{k}} \hbar ck a_{\mathbf{k}}^\dagger a_{\mathbf{k}}$,
and their interaction reads
$V_I  = \sum_{j=1}^N \sum_{\mathbf{k}} \hbar g_{\kbold}
a_\kbold^\dagger |g\rangle\langle e_j| e^{-i\kbold \cdot \rjbold}
e^{i(ck-\omega_0)t}$,
where $g_\kbold$ is the atom-photon coupling constant. The initial state $\Psi_0$
evolves into a state on the form
\begin{equation}\label{eq:psit}
\ket{\Psi(t)} = \sum_{j=1}^N \alpha_j
e^{-i\omega_0t}\ket{e_j}\otimes \ket{0} + \sum_{\mathbf{k}}
\kappa_{\mathbf{k}} e^{-ickt} \ket{g} \otimes \ket{\mathbf{k}},
\end{equation}
where $|g\rangle$ is shorthand for the collective state with all
atoms in the ground state, and where $\alpha_j$ and
$\kappa_{\mathbf{k}}$ are time-dependent expansion coefficients in
the interaction picture. From the formal solution of the
Schr\"{o}dinger equation for the photon state amplitudes
$\kappa_\kbold$, we obtain the atomic amplitude equations
\begin{equation*}
\dot{\alpha}_j = - \sum_{j=1}^N \sum_\kbold |g_\kbold|^2
e^{i\kbold\cdot(\rjbold-\mathbf{r}_{j'})} \int_0^t
e^{i(ck-\omega_0)(t'-t)}\alpha_j(t')dt'.
\end{equation*}
Following \cite{MK07}, we apply the Weisskopf-Wigner approximation and we discard a multi-atom "Lamb-shift" term, which is expected to be at least an order of magnitude smaller than the terms retained in our derivation. Introducing new variables $\beta_j = e^{-i\mathbf{k}_0 \cdot \mathbf{r}_j} \alpha_j$ we arrive, for $ct$ larger than the sample size, at the linear set of equations
\begin{equation}\label{eq:bdot}
\dot{\beta}_j = -\gamma_1 \sum_{j'=1}^N F(\rjbold-\rjpbold)
\beta_{j'},
\end{equation}
where
$\gamma_1= \int \frac{d\Omega_{\mathbf{n}}}{4\pi} \pi |g_\kbold|^2
\rho(ck)|_{\kbold = \mathbf{n}k_0}$, and $2\gamma_1$ is the usual single atom decay rate, and
where
$F(\rjbold-\rjpbold) =
\frac{\sin(k_0|\rjbold-\rjpbold|)}{k_0|\rjbold-\rjpbold|}
e^{-i\mathbf{k}_0 \cdot(\rjbold-\rjpbold)}$.
The fully symmetric, superradiant state (1) decays with the rate,
\begin{equation}\label{eq:gcol}
\gamma_{col} = \frac{\gamma_1}{N}\sum_{j=1}^N \sum_{j'=1}^N
F(\rjbold-\rjpbold),
\end{equation}
but it is not an exact eigenvector for Eq.(\ref{eq:bdot}), which is, however, easily solved for our system with only few hundred atoms by diagonalization of the matrix $F$.

The Schr\"odinger equation for the $\kappa_\kbold$ amplitudes are
first order equations with atomic amplitude source terms, and the
field emanating from the sample is given explicitly by analytical
integrals over time of the exponentially damped atomic eigenmodes of
Eq. \ref{eq:bdot}, weighted by expansion coefficients that we find
from the numerical diagonalization of $F$. In our numerical
simulations we have studied a cubic lattice with an elongated sample
of $7\times 7\times 20$ atoms. With a lattice spacing of 0.37 \mum
$,$ the maximum distance between any two atoms is 8.3 \mum, short
enough to achieve the Rydberg blockade. We use numbers
characteristic for \Rb87 and the 5$P_{1/2}$ excited state with a
spontaneous emission rate of $2\gamma_1 = 37$ $\mu$s$^{-1}$.

As demonstrated in Fig. 1, the excited state population initially
decays as $\exp(-2\gamma_{col}t)$ (dashed line), where $\gamma_{col}=5.7 \gamma_1$. At later times, the symmetry of the atomic excited state population in the sample is broken (see insert in Fig.1),
explaining the longer survival of a few per cent of
collective excitation in the system.

\begin{figure}[tb]
\centering {\includegraphics[width=7cm]{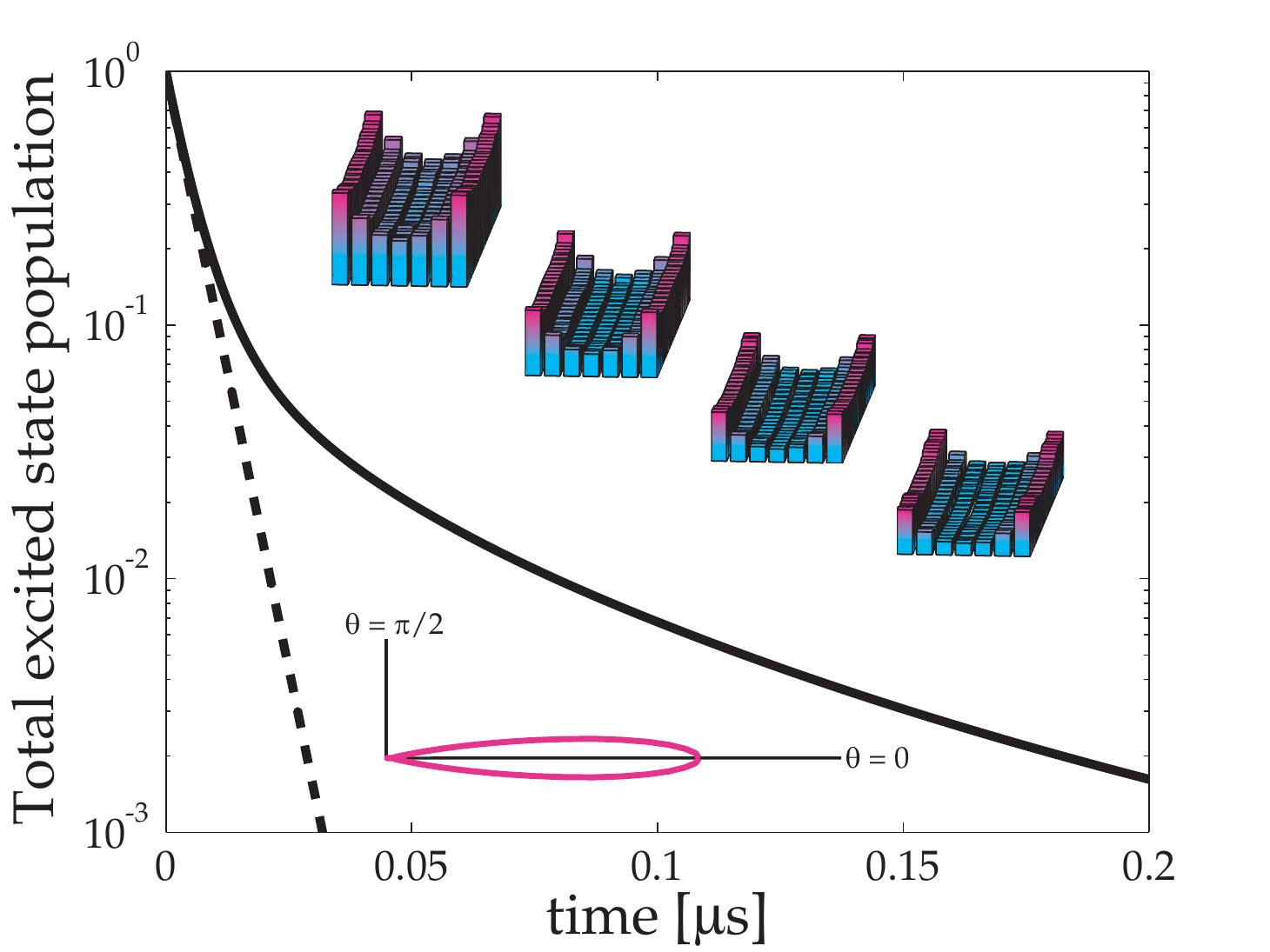}
\caption{(Color online) Excited state population in an $^{87}$Rb
sample with $7\times 7\times 20$ atoms (solid curve). The population
departs from  the exponential collective decay law (dashed line)
around $t=10^{-8}$ s, where the spatial distribution of the excited
state population in the four top layers of the sample is shown in
the upper part of the graph. The bottom insert shows the directional
photon density at $t=10^{-7}$ s.}\label{fig:excitedpop}}
\end{figure}

Phase matching ensures that in the superradiant stage the light
emission occurs predominantly within a narrow emission cone. This is
illustrated in the lower part of Fig. 1, showing the total photon
emission probability as function of direction. With more than 95 \%
probability the photon is emitted in a direction within 0.3 radians
off the axis of the sample.

Having thus established the light emission properties of the
atomic samples, we return to the issue of entanglement generation
and communication between distinct samples.  If the fields emanating
from two synchronously prepared samples are mixed on a beam
splitter, entanglement of such samples is heralded by the detector
outcome \cite{monroe2}. Our architecture provides a high success
probability of this detection scheme due to the superradiant
emission of light into preferred directions and a resulting good
mode matching of the two field modes.

\begin{figure}[t]
\centering {\includegraphics[width=7cm]{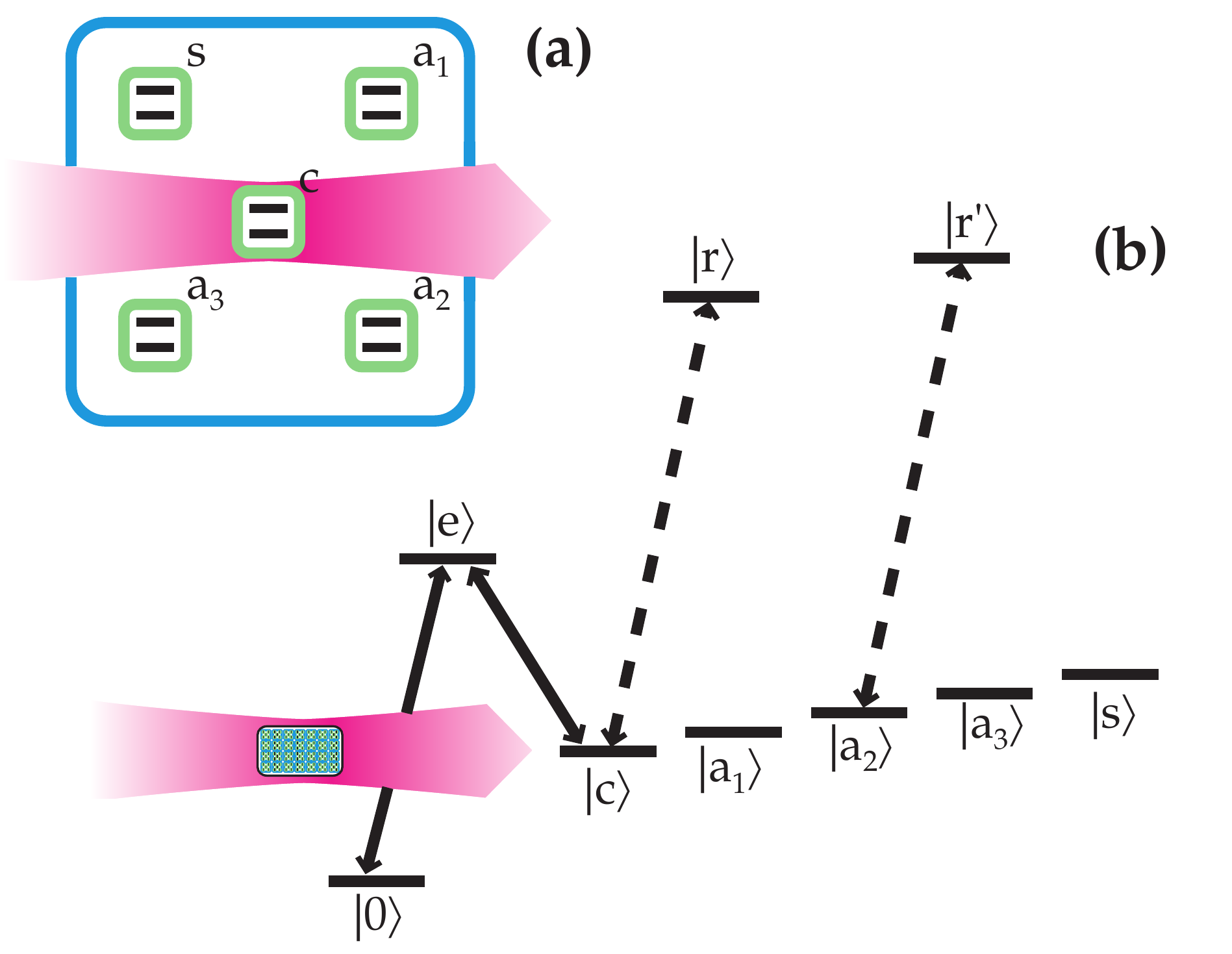}}
\caption{(Color online) (a) A five-qubit register consisting of a
communication qubit ($c$), a storage qubit ($s$) and three auxiliary
qubits ($a_{1,2,3}$) \cite{jiangshort}. (b) The collective encoding
implementation, with a collective internal state transition
interacting with the field mode, and long lived and Rydberg internal
states used for encoding and coupling of the five
qubits.}\label{fig:qubits}
\end{figure}

More than 95 \% of the light is emitted within a well confined mode,
and time reversal ensures that a field with a spatial dependence
which is the complex conjugate of the fields found above will travel
in the opposite direction and become extinct by the atomic
excitation with a probability exceeding 95 \%. This offers the
possibility to split a single photon by a beam splitter and direct
it towards two separate ensembles in a deterministic protocol for
entanglement generation, and if reshaping of single photon pulses
between the mode emitted and its complex conjugate is possible, a photon
wave packet generated in one sample can be absorbed with high
fidelity in another sample in a high fidelity quantum information
transfer protocol.

In the above analysis we have found that a few hundred atoms within a 10 $\mu$m wide volume in space has a high coupling fidelity (in excess of 95 \%) to a single photonic qubit. Using the Rydberg blockade gates \cite{jaksch}, it also offers efficient means for the ensuing entanglement pumping \cite{briegel98} via two-bit gates from the information receiving qubit towards other bits in the register, cf. Fig. 2. The entanglement pumping protocol \cite{jiangshort}, involving auxiliary qubits, measurements and multiple rounds of communication can raise a $90 \%$ transmission fidelity to arbitrarily high degrees of entanglement between two samples. We refer the reader to \cite{jiangshort} for the algorithmic details of the protocol, and turn to the description of our physical implementation of local five-qubit registers needed for the protocol and of their coupling to light.

An ensemble of $K>N$ identical, collectively addressed particles with $(N+1)$  internal levels, $|i\rangle\ {i=0, ... N}$, can encode the qubits of an
$N$-bit register \cite{coll1,coll2}. The $\left \vert i=0 \right \rangle$ state is our "reservoir" state, populated initially by all members of the ensemble, and we  associate the computational register
state $\left \vert b_1,b_2, ..., b_N \right \rangle$, $(b_i =0,1)$ with the \emph{symmetric} state of
the ensemble with $b_i$ ensemble members populating the single atom states $|i\rangle$. To be specific, we shall choose the states $|i\rangle$ as Zeeman sub-levels of the atomic hyperfine ground states of atoms, which we assume to be trapped in a small optical lattice. The register state $|0_1,0_2, ... 0_N\rangle = \bigotimes_{j=1}^K |0\rangle_j$ is the starting point for our analysis.

In a recent publication \cite{saffman08} we have suggested that up to 1000 bit registers may be built, using the collective encoding in holmium atoms, and using multiple individually addressable ensembles which are within the Rydberg interaction distance of each other. In the present work, our interest is in small registers with the capacity to store only five qubits. In the holmium ground state with hyperfine quantum numbers of $F=4, .. ,11$, we have access to 8 different field insensitive $M_F=0$ states, which would make long coherence times of a five-bit register feasible, while a modified storage scheme with qubit values zero and unity encoded in the collective population of state pairs  $\{ |F,M_F\rangle, |F',-M_F\rangle\}$ provides an adequate number of (first order) field insensitive qubits in, e.g., rubidium or cesium \cite{treutlein}.

Figure 2 illustrates the five-qubit register design, proposed in \cite{jiangshort}, and the proposal of the present Letter. In part a) of the figure, five separate physical systems take the role of a communication qubit, "c", three auxiliary qubits for temporary storage and entanglement pumping, "$a_i, i=1,2,3$", and  a storage qubit for the perfected state, "s". A chain of trapped ions with a single ion residing in an optical cavity for communication, or $^{13}$C atoms in the proximity of an optically addressable NV center in diamond are proposed in \cite{jiangshort} as candidates for these five physical qubits.  In part b) of the figure is shown a generic single-atom level scheme for our collective encoding with a reservoir state, and five different long lived states playing the same roles as the five physical qubits in part a) of the figure, and two Rydberg excited states, needed for initialization and one- and two-bit operations. To encode and controllably manipulate an $N$-bit quantum register in a single
mesoscopic ensemble of atoms, one uses the fact that the symmetric collective state of the ensemble with one atom populating a given register state $|i\rangle$ can be selectively transferred to a symmetric state with one atom in the Rydberg excited state $|r\rangle$. From here, due to the blockade effect, coherent driving on the $|0\rangle \leftrightarrow |r\rangle$ transition drives a closed two-level transition in the multi-atom sample, and after return of the remaining $|r\rangle$ state contents into the qubit state $|i\rangle$, one has effectively carried out a one-bit operation on the i'th qubit \cite{coll1}. Two-qubit gates involve the excitation of the control qubit internal state into a Rydberg state, followed by the attempted excitation of the target qubit internal state to another Rydberg state, cf. the dashed arrows in Fig. 3 b). After a $2\pi$ Rabi rotation, the target qubit 1-state thus acquires a sign change depending on the initial population (qubit value) of the control qubit. Note that the two-qubit gate is very similar to the one proposed in \cite{jaksch}, except that it does not require access to individual particles, and it makes use of two different Rydberg states in the atoms.

The absorption of a single photon on the $|0\rangle - |e\rangle$
transition, sketched in the figure, causing a collective excitation
of the form of Eq. 1, must be quickly followed by a resonant
transition between the excited state $|e\rangle$ and the long-lived
"communication state" $|c\rangle$, and the stationary communication
qubit thus acquires the state of the incident travelling qubit with
$\geq 95 \%$ fidelity.  From here, our collective Rydberg gates
between the communication qubit and the auxiliary and storage qubits
are used to implement the algorithm proposed in \cite{jiangshort}.

We note that a general quantum register state is a superposition of
five-bit collective states $|b_1,b_2 .. ,b_5\rangle$ with zero or
unit collective occupancy of the atomic states $c,a_1,a_2,a_3,s$,
and all optical transitions occur, due to the linearity of quantum
mechanics, on every component of that superposition. Since the
number of atoms available for a given transition depends on the
occupancy of all the qubit levels, it may attain values ranging from
$K-5$ to $K$ (when all qubits are equal to unity and zero,
respectively), and laser coupling schemes which yield the precise
pulses, irrespective of such variations, must be employed
\cite{coll1,coll2}. In the collective encoding all qubits may
interchangeably take the roles of communicating, auxiliary and
storage qubits, unless polarization and dipole selection rules make
it advantageous to fix these roles from the beginning,
\textit{e.g.}, to ensure that the decay of the excited state during
photon emission puts the atomic excited state population into the
reservoir state and not into any of the qubit encoding states.

In conclusion, collective encoding of few qubits in small ensembles
of atoms offers a promising approach to interfaces of flying and
stationary qubits, and they hold the potential to provide scalable
quantum computing and long distance quantum communication. We
emphasize that the collective encoding both yields the efficient
coupling to single photons and alleviates the need for addressing of
individual atoms. One might worry that the loss or misplacement of a
few atoms from the sample would cause a significant change in the
field mode, and hence unrealistic demands on the ability to trap
atoms would have to be met. We have tested this concern by removing
up to tens of atoms from random locations in our lattice system. We
have then computed the field mode emitted by the modified structure
and determined the overlap of this field mode with the one emitted
by the complete sample. These overlaps are very robust and in excess
of 99 \% in all our simulations. This also implies that the read
out, needed in the entanglement pumping protocol, can be made by
state selective ionization of qubit internal states, removing a
single atom from the sample for each read out of a "1"-result
without affecting the symmetric state of the remaining atoms.
Another, more serious concern is the gradual destruction of the
symmetric collective state of the system, due to the non-perfect
matching with the superradiant mode. Entanglement pumping can
correct some errors, but when the absorption fails we do not only
have a $<$5 \% qubit error: the system may actually leave the
computational subspace of symmetric states. For a sufficiently large
sample, the system is robust against such errors for a limited
amount of time, and methods exist to counter the errors
\cite{coll2}. We suggest, in addition, to frequently restore the
symmetry of the sample by optically pumping the communication qubit
content into the reservoir state. Another way to obtain a renewable
communication qubit in the sample may be to apply a more elaborate
architecture with individually addressable ensembles within the
Rydberg blockade radius of each other or with a mixture of two
different species, contained within the same volume, and where the
Rydberg blockade may also apply between species. One species, used
for communication may then be optically pumped at any time to
maintain the symmetry of the system, needed for the interaction with
the optical field.

This work was supported by ARO-DTO Grant No.
47949PHQC and the European Union Integrated Project
SCALA. Discussions with Anders S\o rensen are gratefully acknowledged.

\end{document}